\documentclass[a4paper,12pt]{article}
\usepackage{amssymb}
\usepackage{amsmath}
\usepackage{epsfig}
\usepackage{subfigure}
\usepackage{graphics}

\usepackage{latexsym}
\usepackage{rotating}
\usepackage{titlesec}

\title{Multi-component AKNS systems}

\author{Metin G\"{u}rses \thanks{gurses@fen.bilkent.edu.tr}\\
{\small Department of Mathematics, Faculty of Science}\\
{\small Bilkent University, 06800 Ankara - Turkey}\\
Asl{\i} Pekcan \thanks{Email:aslipekcan@hacettepe.edu.tr} \\
{\small Department of Mathematics, Faculty of Science} \\
{\small Hacettepe University, 06800 Ankara - Turkey}
}

\date{\nonumber}
\begin{document}
\maketitle
\date{\nonumber}
\newtheorem{thm}{Theorem}[section]
\newtheorem{Le}{Lemma}[section]
\newtheorem{defi}{Definition}[section]
\newtheorem{ex}{Example}[section]
\newtheorem{pro}{Proposition}[section]
\baselineskip 17pt

\numberwithin{equation}{section}

\begin{abstract}
We study two members of the multi-component AKNS hierarchy. These are multi-NLS and multi-MKdV systems. We derive the Hirota bilinear forms of these equations and obtain soliton solutions. We find all possible local and nonlocal reductions of these systems of equations and give a prescription to obtain their soliton solutions. We derive also $(2+1)$-dimensional extensions of the  multi-component AKNS systems.\\

\noindent \textbf{Keywords.} Multi-component AKNS systems, Local and nonlocal reductions, Hirota method, Negative hierarchy.
\end{abstract}

\section{Introduction}

Systems of integrable nonlinear evolution equations arise in many places. The best way to construct them is to write the associated Lax operators in various Lie algebras \cite{gur-oguz}-\cite{fordy} (see \cite{gur-oguz} and the references therein).
The simplest case is the $sl(2,R)$ algebra valued Lax operators with certain eigenvalue dependencies there arise the AKNS equations \cite{akns}. Changing the algebras one obtains many different coupled systems of integrable evolution equations.

The AKNS family of evolutions is also defined by their associated recursion operator  $\mathcal{R}$,
\begin{equation}\label{generalsys}
cu_{t_n}=\mathcal{R}^{n} u_x \,\, \mathrm{where} \,\,  u= \left( \begin{array}{c}
p  \\
q
 \end{array} \right) \,\, \mathrm{i.e.} \,\,  \left( \begin{array}{c}
c p_{t_n}  \\
c q_{t_n}

 \end{array} \right)= \mathcal{R}^{n} \left( \begin{array}{c}
p_x  \\
q_x
 \end{array} \right),
\end{equation}
\noindent for $n=0,1,2,\cdots$, where
\begin{equation}\label{recursion}
 \mathcal{R}=\left( \begin{array}{cc}
-p\partial^{-1}q+\frac{1}{2}\partial & -p\partial^{-1}p  \\
q\partial^{-1}q & q\partial^{-1}p-\frac{1}{2}\partial
 \end{array} \right)
\end{equation}
\noindent and $c$ is an arbitrary constant. Here $\partial$ is the total $x$-derivative operator and $\partial^{-1}=\int^x$ is the standard anti-derivative. First member ($n=1$) of the hierarchy is the standard AKNS system of equations or the nonlinear Schr\"{o}dinger (NLS) system of equations:
\begin{eqnarray}
&& cp_{t}=\frac{1}{2} p_{xx}-p^2 q, \\
&& cq_{t}=-\frac{1}{2} q_{xx}+q^2 p,
\end{eqnarray}
and the second member ($n=2$) of this hierarchy is the modified Korteweg-de Vries (MKdV) system of equations:
\begin{eqnarray}
&& cp_{t}=\frac{1}{4} p_{xxx}-\frac{3}{2} p q p_{x}, \\
&& cq_{t}=\frac{1}{4} q_{xxx}-\frac{3}{2} q p q_{x}.
\end{eqnarray}

In the last decades there have been considerable amount of works communicated on local and nonlocal reductions, and finding multiple soliton solutions of these systems. The NLS system ($n=1$) and its reductions \cite{AbMu1}-\cite{jianke}, the multidimensional versions of NLS and their reductions \cite{fok}-\cite{gerd3}, the MKdV system ($n=2$) and its reductions \cite{AbMu2}, \cite{AbMu3}, \cite{GurPek2}-\cite{ma} have attracted many researchers to obtain interesting soliton solutions via inverse scattering method, Darboux transformations, and Hirota bilinear method. In \cite{origin}, it was proved that nonlocal reductions are special cases of discrete symmetry transformations.

There are also some interests on higher members of the system (\ref{generalsys}) for $n\ge 3$ \cite{fourthNLS1}-\cite{sixthNLS3}. The solution methods, in particular the Hirota method \cite{Hirota} become more difficult or not available for the systems $n \ge 3$. Similar difficulties arise also for the negative AKNS hierarchy in $(2+1)$-dimensions \cite{GurPek4}, \cite{gur-pek02}.

There are various generalizations of the AKNS system. Among them the one introduced by Ma and Zhou \cite{ma-zhou}, \cite{ma0} is very interesting.
This system is given as
\begin{align*}
& c\alpha {\bf p}_{t}+{\bf p}_{xx}+2 Q {\bf p}=0,\\
& c\alpha {\bf q}_{t}-{\bf q}_{xx}-2 Q {\bf q}=0,
\end{align*}
where both $\alpha$ and $c$ are arbitrary constants, ${\bf p}$ and ${\bf q}$ are $N$-vectors. Here  $Q={\bf p} \cdot {\bf q}$.
 Bi-Hamiltonian structures and Riemann-Hilbert problem have been already studied in \cite{ma0}.

Although we shall introduce these $(1+1)$-dimensional multi-component AKNS systems (multi-AKNS system) of Ma and Zhou in the next section, we display here the first two members of this hierarchy for $N=2$ (four fields) to show why such systems are called multi-NLS and multi-MKdV systems. The multi-NLS system for $N=2$ is
\begin{align}\displaystyle
&c\alpha q_t+q_{xx}+2Qq=0,\label{4NLS-1a}\\
&c\alpha p_t+p_{xx}+2Qp=0,\label{4NLS-1b}\\
&c\alpha r_t-r_{xx}-2Qr=0,\label{4NLS-1c}\\
&c\alpha s_t-s_{xx}-2Qs=0,\label{4NLS-1d}
\end{align}
where $Q=qr+ps$. Indeed in \cite{ma0}, Ma constructed the multi-component
AKNS hierarchy and the above system is the first member of this hierarchy for $N=2$. By the application of the Riemann-Hilbert approach he obtained some soliton solutions of this
integrable multi-component
AKNS hierarchy. In \cite{YWL}, the authors derived a more general multi-component NLS hierarchy consisting the system (\ref{4NLS-1a})-(\ref{4NLS-1d}).

\noindent The multi-MKdV system for $N=2$ is
\begin{align}\displaystyle
&c\alpha^2q_t=q_{xxx}+3 Q q_x+3[q_x r+p_xs]q,\label{4MKdV-1a}\\
&c\alpha^2p_t=p_{xxx}+3 Q p_x+3[q_x r+p_xs]p,\label{4MKdV-1b}\\
&c\alpha^2r_t=r_{xxx}+3 Q r_x+3[r_x q+s_xp]r,\label{4MKdV-1c}\\
&c\alpha^2s_t=s_{xxx}+3 Q s_x+3[r_x q+s_xp]s,\label{4MKdV-1d}
\end{align}
 where $Q=qr+ps$. If $p=q$ and $s=r$ the systems (\ref{4NLS-1a})-(\ref{4NLS-1d}) and (\ref{4MKdV-1a})-(\ref{4MKdV-1d}) reduce to NLS and MKdV systems, respectively. These equations can be defined also by their recursion operators which will be given in the following sections. We derive the Hirota bilinearizations of the first two members, namely the multi-NLS and the multi-MKdV systems of the multi-AKNS hierarchy in the general case. Then we obtain one-soliton solutions of these equations. We discuss the possible local and nonlocal reductions of multi-NLS and multi-MKdV systems for $N=2$ (four fields) and obtain some new systems of nonlocal multi-NLS and multi-MKdV equations and give an approach to obtain soliton solutions of these reduced equations. Finally, we discuss negative multi-AKNS hierarchy. We obtain $(2+1)$-dimensional generalization of the multi-AKNS system first and then present $(2+1)$-extensions of multi-NLS and multi-MKdV equations. We shall communicate the soliton solutions of these equations elsewhere.

\section{Multi-component AKNS systems}
A multi-component AKNS system \cite{ma0} is
\begin{align}
&& c\alpha p_{t}^j+p_{xx}^j+2 (\sum_{l=1}^N p^lq^l) p^{j}=0,\quad j=1,\cdots, N, \label{generalmultia}\\
&& c\alpha q_{t}^j-q_{xx}^j-2 (\sum_{l=1}^N p^lq^l) q^{j}=0,\quad j=1,\cdots, N, \label{generalmultib}
\end{align}
where $p^j=p^j(x,t)$, $q^j=q^j(x,t)$, $j=1,\cdots, N$ are complex dynamical variables in general, $\alpha$ is a real constant, $c$ is an arbitrary constant. Recursion operator of the above system is $\mathcal{R}_N=J_2J_1^{-1}$,
such that $(J_1, J_2)$ is the Hamiltonian pair
given by
\begin{align}\displaystyle
&J_1=\left( \begin{array}{cc}
0 &\alpha I_N \\
-\alpha I_N& 0
 \end{array} \right),\\
 &\nonumber\\
&J_2=\left( \begin{array}{cc}
p^T\partial^{-1}p+(p^T\partial^{-1}p)^T &-(\partial +  \displaystyle \sum_{j=1}^N p^j\partial^{-1}q^j)I_N- p^T\partial^{-1}q^T \\
-(\partial +  \displaystyle \sum_{j=1}^N q^j\partial^{-1}p^j)I_N- q\partial^{-1}p&  q\partial^{-1}p^T+(q\partial^{-1}q^T)^T
 \end{array} \right),
\end{align}
where $I_N$ is the $N\times N$ identity matrix for $N\geq 1$. Here $p=(p_1,\cdots, p_N)$, $q=(q_1,\cdots, q_N)^T$,
and $T$ stands for matrix transpose. We have the hierarchy of multi-component AKNS system written by
\begin{equation}\label{hierarchy}
cU_t=\left(\mathcal{R}_N \right)^n U_x, ~~~~n=1,2,\cdots,\quad U=(p,q^T)^T,
\end{equation}
where
\begin{equation}\label{rec}
\mathcal{R}_{N}=\left( \begin{array}{cc}
R_{1} & R_{2} \\
R_{3}& R_{4}
\end{array} \right),
\end{equation}
for $R_1, R_2, R_3$, and $R_4$ are $N\times N$ matrices which are given by
\begin{eqnarray}
&&\left(R_{1}\right)^{i}_{j}=-\frac{1}{\alpha}\, \left(\partial+p_{k}\, \partial^{-1}\, q_{k} \right)\, \delta^{i}_{j}-\frac{1}{\alpha}\, p^{i}\, \partial^{-1}\, q_{j}, \\
&&\left(R_{2}\right)^{i}_{j}=-\frac{1}{\alpha}\, \left(p^{i} \partial^{-1}\, p_{j} +p_{j} \, \partial^{-1}\, p^{i}\,\right), \\
&&\left(R_{3}\right)^{i}_{j}=\frac{1}{\alpha}\, \left(q^{i} \partial^{-1}\, q_{j} +q_{j} \, \partial^{-1}\, q^{i}\,\right), \\
&&\left(R_{4}\right)^{i}_{j}=\frac{1}{\alpha}\, \left(\partial+q_{k}\, \partial^{-1}\, p_{k} \right)\, \delta^{i}_{j}+\frac{1}{\alpha}\, q^{i}\, \partial^{-1}\, p_{j},
\end{eqnarray}
for $i,j=1,2,\cdots, N$ and $\delta^{i}_{j}$ is the Kronecker delta symbol. Then the evolution equations are obtained from
\begin{eqnarray}
&&cp^{i}_{t}=\left(R_{1}\right)^{i}_{j}\,p^{j}_{x}+\left(R_{2}\right)^{i}_{j}\,q^{j}_{x}, ~~~i=1,2,\cdots, N,\\
&&cq^{i}_{t}=\left(R_{3}\right)^{i}_{j}\,p^{j}_{x}+\left(R_{4}\right)^{i}_{j}\,q^{j}_{x}, ~~~i=1,2,\cdots, N.
\end{eqnarray}
Here we have used the Einstein convention where repeated indices are summed up from 1 to $N$. For $N=2$  the above system reduces to
\begin{align}\displaystyle
&c\alpha p_t^1+p_{xx}^1+2Qp^1=0,\label{4m-NLS-a}\\
&c\alpha p_t^2+p_{xx}^2+2Qp^2=0,\label{4m-NLS-b}\\
&c\alpha q_t^1-q_{xx}^1-2Qq^1=0,\label{4m-NLS-c}\\
&c\alpha q_t^2-q_{xx}^2-2Qq^2=0,\label{4m-NLS-d}
\end{align}
where $Q=p^1q^1+p^2q^2$. The second member ($n=2$) of the hierarchy of the multi-component AKNS system is the multi-MKdV system which is given by
\begin{align}
c  p^{i}_{t}&=-\frac{1}{\alpha}\left(R_{1}\right)^{i}_{j}\,(p^j_{xx}+2 Q p^{j}) +\frac{1}{\alpha}\left(R_{2}\right)^{i}_{j}\,(q^{j}_{xx}+2 Q q^{j}),\quad i=1,2,\cdots, N,\nonumber\\
&=\frac{1}{\alpha^2}(p^{i}_{xxx} + 3Q p^{i}_{x}+3 (q_{k}\,p^{k}_{x})\,p^{i}),\label{multiMKdVa}\\
c q^{i}_{t}&=-\frac{1}{\alpha}\left(R_{3}\right)^{i}_{j}(p^j_{xx}+2 Q p^{j})+\frac{1}{\alpha}\left(R_{4}\right)^{i}_{j}\,(q^{j}_{xx}+2 Q q^{j}),\quad i=1,2,\cdots, N,\nonumber\\
&=\frac{1}{\alpha^2}(q^{i}_{xxx}+3 Q q^{i}_{x}+3 (p_{k}\,q^{k}_{x})\,q^{i}),\label{multiMKdVb}
\end{align}
where $ \displaystyle Q=\sum_{l=1}^N p^lq^l$. For $N=2$, the above system becomes
\begin{align}
&c\alpha^2p_t^1=p_{xxx}^1+3Q\,p_x^1+3[p_x^1q_1+p_x^2q^2]p^1,\\
&c\alpha^2p_t^2=p_{xxx}^2+3Q\,p_x^2+3[p_x^1q_1+p_x^2q^2]p^2,\\
&c\alpha^2q_t^1=q_{xxx}^1+3Q\,q_x^1+3[q_x^1p_1+q_x^2p^2]q^1,\\
&c\alpha^2q_t^2=q_{xxx}^1+3Q\,q_x^2+3[q_x^1p_1+q_x^2p^2]q^2.
\end{align}

In the next section we mainly focus on multi-NLS and multi-MKdV equations for $N=2$ (four fields). These equations are simple generalizations of the NLS and MKdV systems. Hence we study the Hirota bilinearization of these systems of equations.

\section{Hirota bilinear forms of the multi-component AKNS systems}

For the multi-component AKNS system (\ref{generalmultia}) and
(\ref{generalmultib}) which is the first member $(n=1)$ of the hierarchy (\ref{hierarchy}) that we call as the multi-NLS system, by letting $\displaystyle p^j=\frac{g^j}{f}$ and $\displaystyle q^j=\frac{h^j}{f}$, for $j=1,\cdots, N$ we get the Hirota bilinear form as
\begin{align}
&(c\alpha D_t+D_x^2)\{g^j\cdot f\}=0,\quad j=1,\cdots, N,\label{MaZhouHa}\\
&(c\alpha D_t-D_x^2)\{h^j\cdot f\}=0,\quad j=1,\cdots, N,\label{MaZhouHb}\\
&D_x^2\{f\cdot f\}=2 \sum_{l=1}^Ng^lh^l.\label{MaZhouHc}
\end{align}

\noindent For the multi-MKdV system that is the second member $(n=2)$ of the hierarchy (\ref{hierarchy}), letting $\displaystyle p^j=\frac{g^j}{f}$ and $\displaystyle q^j=\frac{h^j}{f}$, for $j=1,\cdots, N$ gives the Hirota bilinear form of (\ref{multiMKdVa}) and (\ref{multiMKdVb}) as
\begin{align}{\displaystyle}
&(c\alpha^2 D_t-D_x^3)\{g^j\cdot f\}+3\tau^j=0,\quad j=1,\cdots, N,\label{N-MKdV-Ha}\\
&(c\alpha^2 D_t-D_x^3)\{h^j\cdot f\}+3\mu^j=0,\quad j=1,\cdots, N,\label{N-MKdV-Hb}\\
&\sum_{m \ne j}^{N} h^mD_x\{g^j\cdot g^m\} =\tau^j f,\quad  j=1,\cdots, N,\label{N-MKdV-Hc}\\
&\sum_{m \ne j}^{N} g^mD_x\{h^j\cdot h^m\} =\mu^j f,\quad  j=1,\cdots, N, \label{N-MKdV-Hd}\\
&D_x^2\{f\cdot f\}=2 \sum_{l=1}^Ng^lh^l,\label{N-MKdV-He}
\end{align}
where $\tau^j$ and $\mu^j$, $j=1,\cdots, N$, are auxiliary functions of $(x,t)\in \mathbb{R}^2$.

\subsection{Soliton solutions of the multi-NLS system for $N=2$}

From (\ref{MaZhouHa})-(\ref{MaZhouHc}) we get the Hirota bilinear form of the multi-NLS system for $N=2$ as
\begin{align}
&(c\alpha D_t+D_x^2)\{g^1\cdot f\}=0,\label{H-a}\\
&(c\alpha D_t+D_x^2)\{g^2\cdot f\}=0,\label{H-b}\\
&(c\alpha D_t-D_x^2)\{h^1\cdot f\}=0,\label{H-c}\\
&(c\alpha D_t-D_x^2)\{h^2\cdot f\}=0,\label{H-d}\\
&D_x^2\{f\cdot f\}=2(g^1h^1+g^2h^2).\label{H-e}
\end{align}
Similarly, the Hirota bilinear form of the multi-MKdV system for $N=2$ is obtained from (\ref{N-MKdV-Ha})-(\ref{N-MKdV-He}) as
\begin{align}
&(c\alpha^2 D_t-D_x^3)\{g^1\cdot f\}+3\tau=0,\label{HMKdVN=2-a}\\
&(c\alpha^2D_t-D_x^3)\{g^2\cdot f\}+3\rho=0,\label{HMKdVN=2-b}\\
&(c\alpha^2 D_t-D_x^3)\{h^1\cdot f\}+3\mu=0,\label{HMKdVN=2-c}\\
&(c\alpha^2D_t-D_x^3)\{h^2\cdot f\}+3\sigma=0,\label{HMKdVN=2-d}\\
&h^2D_x\{g^1\cdot g^2\}=\tau f,\label{HMKdVN=2-e}\\
&h^1D_x\{g^2\cdot g^1\}=\rho f,\label{HMKdVN=2-f}\\
&g^2D_x\{h^1\cdot h^2\}=\mu f,\label{HMKdVN=2-g}\\
&g^1D_x\{h^2\cdot h^1 \}=\sigma f,\label{HMKdVN=2-h}\\
&D_x^2\{f\cdot f\}=2(g^1h^1+g^2h^2),\label{HMKdVN=2-i}
\end{align}
where $\tau, \rho, \mu,$ and $\sigma$ are auxiliary functions of $(x,t)\in \mathbb{R}^2 $.

\subsubsection{One-soliton solutions of the multi-NLS system for $N=2$}

Use the following expansion in the Hirota bilinear form (\ref{H-a})-(\ref{H-e});
\begin{align}\displaystyle
&g^1=\varepsilon g_1^1+\varepsilon^3 g_3^1,\quad g^1=\varepsilon g_1^1+\varepsilon^3 g_3^1,\quad h^1=\varepsilon h_1^1+\varepsilon^3 h_3^1,\\
&h^2=\varepsilon h_1^2+\varepsilon^3 h_3^2,\quad f=1+\varepsilon^2 f_2+\varepsilon^4 f_4,
\end{align}
where
\begin{equation}\displaystyle
g^1=e^{\theta_1},\quad g^2=e^{\theta_2},\quad h^1=e^{\theta_3},\quad h^2=e^{\theta_4}
\end{equation}
for $\theta_j=k_jx+\omega_j t+\delta_j$, $j=1,2,3,4$. We make the coefficients of $\varepsilon^m$, $m=1,2,\cdots, 8$ vanish. The coefficients of
$\varepsilon$ give the dispersion relations
\begin{equation}\label{dispersion}\displaystyle
\omega_1=-\frac{1}{c\alpha}k_1^2,\quad \omega_2=-\frac{1}{c\alpha}k_2^2,\quad \omega_3=\frac{1}{c\alpha}k_3^2,\quad \omega_4=\frac{1}{c\alpha}k_4^2.
\end{equation}
From the coefficient of $\varepsilon^2$ we get
\begin{equation}\displaystyle
f_2=\Big[\frac{e^{\theta_1+\theta_3}}{(k_1+k_3)^2}+\frac{e^{\theta_2+\theta_4}}{(k_2+k_4)^2}\Big].
\end{equation}
The coefficients of $\varepsilon^3$ give
\begin{align}
&g_3^1=\gamma_1 e^{\theta_1+\theta_2+\theta_4},\quad g_3^2=\gamma_2 e^{\theta_1+\theta_2+\theta_3},\\
&h_3^1=\gamma_3 e^{\theta_2+\theta_3+\theta_4},\quad h_3^2=\gamma_4 e^{\theta_1+\theta_3+\theta_4},
\end{align}
where
\begin{align}\label{gammalar}\displaystyle
&\gamma_1=\frac{(k_1-k_2)}{(k_1+k_4)(k_2+k_4)^2}, \quad \gamma_2=-\frac{(k_1-k_2)}{(k_2+k_3)(k_1+k_3)^2},\\
&\gamma_3=\frac{(k_3-k_4)}{(k_2+k_3)(k_2+k_4)^2}, \quad \gamma_4=-\frac{(k_3-k_4)}{(k_1+k_4)(k_1+k_3)^2}.
\end{align}
From the coefficient of $\varepsilon^4$ we obtain the function $f_4$ as
\begin{equation}
f_4=Me^{\theta_1+\theta_2+\theta_3+\theta_4},
\end{equation}
where
\begin{equation}\label{M}\displaystyle
M=\frac{(k_1-k_2)(k_3-k_4)}{(k_1+k_4)(k_2+k_3)(k_1+k_3)^2(k_2+k_4)^2}.
\end{equation}
The coefficients of $\varepsilon^5$, $\varepsilon^6$, $\varepsilon^7$, and $\varepsilon^8$ vanish directly. Without loss of generality take $\varepsilon=1$.
Hence one-soliton solution of the multi-NLS system (\ref{generalmultia}) and (\ref{generalmultib}) for $N=2$ is given by $(p^1,p^2,q^1,q^2)$ where
\begin{align}\displaystyle
&p^1=\frac{g^1}{f}=\frac{e^{\theta_1}+\gamma_1e^{\theta_1+\theta_2+\theta_4}}{1+\Big[\frac{e^{\theta_1+\theta_3}}{(k_1+k_3)^2}+\frac{e^{\theta_2+\theta_4}}{(k_2+k_4)^2}
\Big]+Me^{\theta_1+\theta_2+\theta_3+\theta_4}}, \label{sol1}\\
&p^2=\frac{g^2}{f}=\frac{e^{\theta_2}+\gamma_2e^{\theta_1+\theta_2+\theta_3}}{1+\Big[\frac{e^{\theta_1+\theta_3}}{(k_1+k_3)^2}+\frac{e^{\theta_2+\theta_4}}{(k_2+k_4)^2}
\Big]+Me^{\theta_1+\theta_2+\theta_3+\theta_4}},\label{sol2}\\
&q^1=\frac{h^1}{f}=\frac{e^{\theta_3}+\gamma_3e^{\theta_2+\theta_3+\theta_4}}{1+\Big[\frac{e^{\theta_1+\theta_3}}{(k_1+k_3)^2}+\frac{e^{\theta_2+\theta_4}}{(k_2+k_4)^2}
\Big]+Me^{\theta_1+\theta_2+\theta_3+\theta_4}}, \label{sol3}\\
&q^2=\frac{h^2}{f}=\frac{e^{\theta_4}+\gamma_4e^{\theta_1+\theta_3+\theta_4}}{1+\Big[\frac{e^{\theta_1+\theta_3}}{(k_1+k_3)^2}+\frac{e^{\theta_2+\theta_4}}{(k_2+k_4)^2}
\Big]+Me^{\theta_1+\theta_2+\theta_3+\theta_4}}, \label{sol4}
\end{align}
with $\theta_j=k_jx+\omega_j t+\delta_j$, $j=1,2,3,4$ satisfying the dispersion relations (\ref{dispersion}).

\subsubsection{One-soliton solutions of the multi-MKdV system for $N=2$}

Use the following expansion in the Hirota bilinear form (\ref{HMKdVN=2-a})-(\ref{HMKdVN=2-i});
\begin{align}\displaystyle
&g^1=\varepsilon g_1^1+\varepsilon^3 g_3^1,\quad g^1=\varepsilon g_1^1+\varepsilon^3 g_3^1,\quad h^1=\varepsilon h_1^1+\varepsilon^3 h_3^1,\\
&h^2=\varepsilon h_1^2+\varepsilon^3 h_3^2,\quad f=1+\varepsilon^2 f_2+\varepsilon^4 f_4, \quad \tau=\varepsilon^3\tau_3+\varepsilon^5\tau_5,\\
&\rho=\varepsilon^3\rho_3+\varepsilon^5\rho_5,\quad \mu=\varepsilon^3\mu_3+\varepsilon^5\mu_5,\quad \sigma=\varepsilon^3\sigma_3+\varepsilon^5\sigma_5,
\end{align}
where
\begin{equation}\displaystyle
g^1=e^{\theta_1},\quad g^2=e^{\theta_2},\quad h^1=e^{\theta_3},\quad h^2=e^{\theta_4}
\end{equation}
for $\theta_j=k_jx+\omega_j t+\delta_j$, $j=1,2,3,4$. We make the coefficients of $\varepsilon^m$, $m=1,2,\cdots, 9$ vanish. The coefficients of
$\varepsilon$ give the dispersion relations
\begin{equation}\label{dispersionMKdV}\displaystyle
\omega_j=\frac{1}{c\alpha^2}k_j^3,\quad j=1, 2, 3, 4.
\end{equation}
From the coefficient of $\varepsilon^2$ we get
\begin{equation}\displaystyle
f_2=\Big[\frac{e^{\theta_1+\theta_3}}{(k_1+k_3)^2}+\frac{e^{\theta_2+\theta_4}}{(k_2+k_4)^2}\Big].
\end{equation}
The coefficients of $\varepsilon^3$ give
\begin{align}
&g_3^1=\gamma_1 e^{\theta_1+\theta_2+\theta_4},\quad g_3^2=\gamma_2 e^{\theta_1+\theta_2+\theta_3},\\
&h_3^1=\gamma_3 e^{\theta_2+\theta_3+\theta_4},\quad h_3^2=\gamma_4 e^{\theta_1+\theta_3+\theta_4},\\
&\tau_3=(k_1-k_2)e^{\theta_1+\theta_2+\theta_4},\quad \rho_3=(k_1-k_2)e^{\theta_1+\theta_2+\theta_3},\\
&\mu_3=(k_3-k_4)e^{\theta_2+\theta_3+\theta_4},\quad \sigma_3=(k_3-k_4)e^{\theta_1+\theta_3+\theta_4},
\end{align}
where the functions $\gamma_j, j=1, 2, 3, 4$ are the same as in (\ref{gammalar}). From the coefficient of $\varepsilon^4$ we obtain the function $f_4$ as
\begin{equation}
f_4=Me^{\theta_1+\theta_2+\theta_3+\theta_4},
\end{equation}
where $M$ is same with (\ref{M}). The coefficients of $\varepsilon^5$ give the functions $\tau_5, \rho_5, \mu_5,$ and $\sigma_5$ as
\begin{align}\displaystyle
&\tau_5=-\frac{(k_1-k_2)(k_3-k_4)}{(k_1+k_4)(k_1+k_3)^2}e^{2\theta_1+\theta_2+\theta_3+\theta_4},\quad \rho_5=\frac{(k_1-k_2)(k_3-k_4)}{(k_2+k_3)(k_2+k_4)^2}e^{\theta_1+2\theta_2+\theta_3+\theta_4},\\
&\mu_5=-\frac{(k_1-k_2)(k_3-k_4)}{(k_2+k_3)(k_1+k_3)^2}e^{\theta_1+\theta_2+2\theta_3+\theta_4}, \quad \sigma_5=\frac{(k_1-k_2)(k_3-k_4)}{(k_1+k_4)(k_2+k_4)^2}e^{\theta_1+\theta_2+\theta_3+2\theta_4}.
\end{align}
The coefficients of $\varepsilon^6$, $\varepsilon^7$, $\varepsilon^8$, and $\varepsilon^9$ vanish directly. Take $\varepsilon=1$.
Hence one-soliton solution of the multi-MKdV system for $N=2$ is exactly same with the multi-NLS system for $N=2$ except the dispersion relations (\ref{dispersionMKdV}) as in the case of coupled NLS and MKdV systems for $N=1$ given in \cite{superpos}.

Here we have given only one-soliton solutions of multi-NLS and multi-KdV systems by using the Hirota method. We indeed have two-soliton solutions of these systems of equations but due to their lengths we are not presenting them here.

\section{Reductions}

For a given integrable system of nonlinear evolution equations we have mainly two kinds of reductions; local and nonlocal reductions. The reduced equations are also integrable and admit soliton solutions when the reductions are done consistently.

\subsection{Local reductions}

There are various kinds of reductions for the multi-component AKNS system. We shall discuss the reductions of this system when $N=2$. For $N=2$ the multi-NLS system  is given by
\begin{equation}
c\left( \begin{array}{c}
p^1 \\
p^2\\
q^1\\
q^2
 \end{array} \right)_t=\mathcal{R}_2^n \left( \begin{array}{c}
p^1 \\
p^2\\
q^1\\
q^2
 \end{array} \right)_x.
\end{equation}

\noindent If we let $n=1$, $p_1=q, p_2=p, q_1=r, q_2=s$ we get the following four-component NLS system \cite{ma-zhou}
\begin{align}\displaystyle
&c\alpha q_t+q_{xx}+2Qq=0,\label{4NLS-a}\\
&c\alpha p_t+p_{xx}+2Qp=0,\label{4NLS-b}\\
&c\alpha r_t-r_{xx}-2Qr=0,\label{4NLS-c}\\
&c\alpha s_t-s_{xx}-2Qs=0,\label{4NLS-d}
\end{align}
where $Q=qr+ps$.

\noindent Furthermore the multi-MKdV system is given by
\begin{align}\displaystyle
&c\alpha^2q_t=q_{xxx}+3Qq_x+3[q_xr+p_xs]q,\label{4MKdV-a}\\
&c\alpha^2p_t=p_{xxx}+3Qp_x+3[q_xr+p_xs]p,\label{4MKdV-b}\\
&c\alpha^2r_t=r_{xxx}+3Qr_x+3[r_xq+s_xp]r,\label{4MKdV-c}\\
&c\alpha^2s_t=s_{xxx}+3Qs_x+3[r_xq+s_xp]s.\label{4MKdV-d}
\end{align}

We obtain several new integrable systems of equations from the above systems by employing local and nonlocal reductions. The local reductions are of four kinds.\\

\noindent \textbf{1)} $p=\rho_1 q$, $s=\rho_2 r$, where $\rho_1, \rho_2$ are arbitrary constants.

\noindent Then the multi-NLS system (\ref{4NLS-a})-(\ref{4NLS-d}) reduces to
\begin{eqnarray}\displaystyle
&c \alpha q_t+q_{xx}+2\tilde{Q}q=0, \label{non1}\\
&c \alpha r_t-r_{xx}-2\tilde{Q}r=0, \label{non2}
\end{eqnarray}
\noindent where $\tilde{Q}=(1+\rho_{1}\,\rho_{2}) q r$. By choosing $\rho_{1}\,\rho_{2}=-1$ the above system decouples and reduces to linear heat equations. Indeed such a limit exists for any $N$. Similarly with the same reduction, the multi-MKdV system (\ref{4MKdV-a})-(\ref{4MKdV-d}) reduces to
{\small \begin{align}\displaystyle
&c\alpha^2q_t-q_{xxx}-3\tilde{Q}q_x-3\rho q_xqr=0,\\
&c\alpha^2r_t-r_{xxx}-3\tilde{Q}r_x-3\rho r_xqr=0,
\end{align}}
\noindent where $\rho=1+\rho_{1} \rho_{2}$. With the special case $\rho_{1}\,\rho_{2}=-1$, multi-MKdV system reduces to two decoupled linear third order differential equations. If $r=\ell q$, for $\ell$ constant, then the above system reduces to a MKdV equation for $q$. If $r$ is taken to be a constant then the above system of equations reduces to the KdV equation.\\

\noindent \textbf{2)}  For the multi-MKdV system (\ref{4MKdV-a})-(\ref{4MKdV-d}) we have a case $s=0$ and $r=r_{0}$ constant, where this system reduces to the following KdV system of equations:
\begin{align}\displaystyle
&c\alpha^2q_t-q_{xxx}-6r_{0}qq_x=0,\\
&c\alpha^2p_t-p_{xxx}-3r_{0}(qp_x+q_xp)=0.
\end{align}
Such a KdV system has been discussed before \cite{Gur-Kar}.\\

\noindent \textbf{3)} $p=\rho_1 \bar{q}$, $s=\rho_2 \bar{r}$, where $\rho_1, \rho_2$ are arbitrary constants. Here bar notation stands for complex conjugation.

\noindent Under this reduction, the multi-NLS system (\ref{4NLS-a})-(\ref{4NLS-d}) reduces consistently to
\begin{align}
&c\alpha q_t+q_{xx}+2(qr+\rho_1\rho_2\bar{q}\bar{r})q=0,\\
&c\alpha r_t-r_{xx}-2(qr+\rho_1\rho_2\bar{q}\bar{r})r=0,
\end{align}
if $c=\bar{c}$ and $\rho_1\rho_2=1$. With the same conditions the multi-MKdV system (\ref{4MKdV-a})-(\ref{4MKdV-d}) reduces to
{\small \begin{align*}
&c\alpha^2 q_t-q_{xxx}-3(qr+\rho_1\rho_2\bar{q}\bar{r})q_x-3(q_xr+\rho_1\rho_2\bar{q}_x\bar{r})q=0,\\
&c\alpha^2 r_t-r_{xxx}-3(qr+\rho_1\rho_2\bar{q}\bar{r})r_x-3(r_xq+\rho_1\rho_2\bar{r}_x\bar{q})r=0.
\end{align*}}

\noindent \textbf{4)} $r=\rho_1 \bar{q}$, $s=\rho_2 \bar{p}$, where $\rho_1, \rho_2$ are arbitrary constants.

\noindent Here the multi-NLS system (\ref{4NLS-a})-(\ref{4NLS-d}) reduces to
\begin{align}
&c\alpha q_t+q_{xx}+2(\rho_1q\bar{q}+\rho_2p\bar{p})q=0,\\
&c\alpha p_t+p_{xx}+2(\rho_1q\bar{q}+\rho_2p\bar{p})p=0,
\end{align}
consistently if $c=-\bar{c}$ and $\rho_1=\bar{\rho}_1$, $\rho_2=\bar{\rho}_2$. By the same reduction the multi-MKdV system (\ref{4MKdV-a})-(\ref{4MKdV-d}) reduces consistently to
{\small \begin{align*}
&c\alpha^2 q_t-q_{xxx}-3(\rho_1q\bar{q}+\rho_2p\bar{p})q_x-3(\rho_1q_x\bar{q}+\rho_2p_x\bar{p})q=0,\\
&c\alpha^2 p_t-p_{xxx}-3(\rho_1q\bar{q}+\rho_2p\bar{p})p_x-3(\rho_1q_x\bar{q}+\rho_2p_x\bar{p})p=0,
\end{align*}}
under the conditions $c=\bar{c}$ and $\rho_1=\bar{\rho}_1$, $\rho_2=\bar{\rho}_2$ hold.

\subsection{Nonlocal reductions}

Similar to local reductions we shall discuss the nonlocal reductions also in the case of $N=2$. What we mention in this special case can be easily extended to all $N$.\\

\noindent \textbf{1)} $p(x,t)=\rho_{1}\, q(\epsilon_{1}\,x, \epsilon_{2} t)$ and $s(x,t)=\rho_{2}\, r(\epsilon_{1}\,x, \epsilon_{2} t)$, where $\rho_1, \rho_2$ are arbitrary constants and $\epsilon_1^2=\epsilon_2^2=1$. Then the reduced nonlocal equations are

\noindent Multi-NLS system:
\begin{align}\displaystyle
&c\alpha q_t(x,t)+q_{xx}(x,t)+2Q^{e}q(x,t)=0, \label{non3}\\
&c\alpha r_t(x,t)-r_{xx}(x,t)-2Q^{e}r(x,t)=0, \label{non4}
\end{align}
\noindent where $Q^{e}=q(x,t)r(x,t)+\rho_1\rho_2q(\epsilon_{1}\,x, \epsilon_{2} t)r(\epsilon_{1}\,x, \epsilon_{2} t)$. For consistency we have $\rho_1\rho_2=1$ and $(\epsilon_1,\epsilon_2)=(\pm 1,1)$. The case $(\epsilon_1,\epsilon_2)=(1,1)$ corresponds to a local reduced multi-NLS system.\\

\noindent Multi-MKdV system:
{\small\begin{align}\displaystyle
&c\alpha^2q_t(x,t)-q_{xxx}(x,t)-3Q^{e}q_x(x,t)-3[q_x (x,t)r(x,t)+\rho_{1}\rho_{2}q_x(\epsilon_{1}x, \epsilon_{2}t)r(\epsilon_{1}x, \epsilon_{2} t)]q(x,t)=0,  \nonumber \\
&c\alpha^2r_t(x,t)-r_{xxx}(x,t)-3Q^{e}r_x(x,t)-3[r_x (x,t)q(x,t)+\rho_{1}\rho_{2}r_x(\epsilon_{1}x, \epsilon_{2}t)q(\epsilon_{1}x, \epsilon_{2} t)]r(x,t)=0, \nonumber
\end{align}}
\noindent where $Q^{e}=q(x,t)r(x,t)+\rho_1\rho_2 q(\epsilon_{1}x, \epsilon_{2}t)r(\epsilon_{1}x, \epsilon_{2} t)$. For consistency we have $\rho_1\rho_2=1$ and $\epsilon_{1}\epsilon_{2}=1$.\\

\noindent \textbf{2)} $p(x,t)=\rho_{1}\, \bar{q}(\epsilon_{1}x, \epsilon_{2}t)$ and $s(x,t)=\rho_{2}\bar{r}(\epsilon_{1}x, \epsilon_{2} t)$, where $\rho_1, \rho_2$ are arbitrary constants and $\epsilon_1^2=\epsilon_2^2=1$.

\noindent By this nonlocal reduction the multi-NLS system (\ref{4NLS-a})-(\ref{4NLS-d}) reduces to
\begin{align*}\displaystyle
&c\alpha q_t(x,t)+q_{xx}(x,t)+2\tilde{Q}^{e}q(x,t)=0,\\
&c\alpha r_t(x,t)-r_{xx}(x,t)-2\tilde{Q}^{e}r(x,t)=0,
\end{align*}
\noindent where $\tilde{Q}^{e}=q(x,t)r(x,t)+\rho_1\rho_2\bar{q}(\epsilon_{1}x, \epsilon_{2} t)\bar{r}(\epsilon_{1}x, \epsilon_{2} t)$. Here for consistency we must have
$\rho_1\rho_2=1$ and $c=\bar{c}\epsilon_2$.

The reduced nonlocal multi-MKdV system is
{\small \begin{align*}\displaystyle
&c\alpha^2q_t(x,t)-q_{xxx}(x,t)-3\tilde{Q}^{e}\,q_x(x,t)-3\, [q_x (x,t)r(x,t)+\rho_{1}\rho_{2}\bar{q}_x (\epsilon_{1}x, \epsilon_{2} t)\bar{r}(\epsilon_{1}x, \epsilon_{2} t)]q(x,t)=0,  \nonumber \\
&c\alpha^2r_t(x,t)-r_{xxx}(x,t)-3\tilde{Q}^{e}\,r_x(x,t)-3\, [r_x (x,t)q(x,t)+\rho_{1}\rho_{2}\bar{r}_x (\epsilon_{1}x, \epsilon_{2} t)\bar{q}(\epsilon_{1}x, \epsilon_{2} t)]r(x,t)=0, \nonumber
\end{align*}}
\noindent where $\tilde{Q}^{e}=q(x,t)r(x,t)+\rho_1\rho_2\bar{q}(\epsilon_{1}x, \epsilon_{2} t)\bar{r}(\epsilon_{1}x, \epsilon_{2} t)$, $\rho_1\rho_2=1$, and $c=\bar{c}\epsilon_1\epsilon_2$.\\

\noindent \textbf{3)} $r(x,t)=\rho_{1}q(\epsilon_{1}x, \epsilon_{2} t)$ and $s(x,t)=\rho_{2}\, p(\epsilon_{1}x, \epsilon_{2} t)$, where $\rho_1, \rho_2$ are arbitrary constants and $\epsilon_1^2=\epsilon_2^2=1$.

\noindent Under this nonlocal reduction the multi-NLS system (\ref{4NLS-a})-(\ref{4NLS-d}) consistently reduces to
\begin{align*}
&c\alpha q_t(x,t)+q_{xx}(x,t)+2(\rho_1q(x,t)q(\epsilon_{1}x, \epsilon_{2} t)+\rho_2p(x,t)p(\epsilon_{1}x, \epsilon_{2}t))q(x,t)=0,\\
&c\alpha p_t(x,t)+p_{xx}(x,t)+2(\rho_1q(x,t)q(\epsilon_{1}x, \epsilon_{2} t)+\rho_2p(x,t)p(\epsilon_{1}x, \epsilon_{2}t))p(x,t)=0,
\end{align*}
if $\epsilon_2=-1$. By the same reduction the multi-MKdV system (\ref{4MKdV-a})-(\ref{4MKdV-d}) reduces consistently to
{\small \begin{align*}
&c\alpha^2 q_t(x,t)-q_{xxx}(x,t)-3W^{\epsilon}q_x(x,t)-3[\rho_1q_x(x,t)q(\epsilon_{1}x, \epsilon_{2}t)+\rho_2p_x(x,t)p(\epsilon_{1}x, \epsilon_{2}t)]q(x,t)=0,\\
&c\alpha^2 p_t(x,t)-p_{xxx}(x,t)-3W^{\epsilon}p_x(x,t)-3[\rho_1q_x(x,t)q(\epsilon_{1}x, \epsilon_{2}t)+\rho_2p_x(x,t)p(\epsilon_{1}x, \epsilon_{2}t)]p(x,t)=0,
\end{align*}}
\noindent where $W^{\epsilon}=\rho_1q(x,t)q(\epsilon_{1}x, \epsilon_{2}t)+\rho_2p(x,t)p(\epsilon_{1}x, \epsilon_{2}t)$ and $\epsilon_1\epsilon_2=1$.\\

\noindent \textbf{4)} $r(x,t)=\rho_{1}\, \bar{q}(\epsilon_{1}x, \epsilon_{2} t)$ and $s(x,t)=\rho_{2}\bar{p}(\epsilon_{1}x, \epsilon_{2}t)$, where $\rho_1, \rho_2$ are arbitrary constants and $\epsilon_1^2=\epsilon_2^2=1$.

\noindent When we use this reduction the multi-NLS system (\ref{4NLS-a})-(\ref{4NLS-d}) consistently reduces to
\begin{align*}
&c\alpha q_t(x,t)+q_{xx}(x,t)+2(\rho_1q(x,t)\bar{q}(\epsilon_{1}\,x, \epsilon_{2} t)+\rho_2p(x,t)\bar{p}(\epsilon_{1}\,x, \epsilon_{2} t))q(x,t)=0,\\
&c\alpha p_t(x,t)+p_{xx}(x,t)+2(\rho_1q(x,t)\bar{q}(\epsilon_{1}\,x, \epsilon_{2} t)+\rho_2p(x,t)\bar{p}(\epsilon_{1}\,x, \epsilon_{2} t))p(x,t)=0,
\end{align*}
if $c=-\bar{c}\epsilon_2$, $\rho_1=\bar{\rho}_1$, and $\rho_2=\bar{\rho}_2$. In \cite{MaHW2021}, Ma et al. studied the case when $(\epsilon_{1}, \epsilon_{2})=(-1,1)$
and gave soliton solutions of this coupled nonlocal NLS system by the help of inverse scattering method.

Under the same reduction the multi-MKdV system (\ref{4MKdV-a})-(\ref{4MKdV-d}) reduces to
{\small\begin{align*}
&c\alpha^2 q_t(x,t)-q_{xxx}(x,t)-3\tilde{W}^{\epsilon}q_x(x,t)-3[\rho_1q_x(x,t)\bar{q}(\epsilon_{1}x, \epsilon_{2}t)+\rho_2p_x(x,t)\bar{p}(\epsilon_{1}x, \epsilon_{2}t)]q(x,t)=0,\\
&c\alpha^2 p_t(x,t)-p_{xxx}(x,t)-3\tilde{W}^{\epsilon}p_x(x,t)-3[\rho_1q_x(x,t)\bar{q}(\epsilon_{1}x, \epsilon_{2}t)+\rho_2p_x(x,t)\bar{p}(\epsilon_{1}x, \epsilon_{2}t)]p(x,t)=0,
\end{align*}}
where $\tilde{W}^{\epsilon}=\rho_1q(x,t)\bar{q}(\epsilon_{1}x, \epsilon_{2}t)+\rho_2p(x,t)\bar{p}(\epsilon_{1}x, \epsilon_{2}t)$, $c=\bar{c}\epsilon_1\epsilon_2$, $\rho_1=\bar{\rho}_1$, and $\rho_2=\bar{\rho}_2$.

\subsection{Soliton solutions of the reduced multi-component AKNS equations}

To obtain soliton solutions of the reduced multi-component AKNS equations we do not need to construct new Hirota bilinear forms or use another known techniques. For this purpose we need only the soliton solutions of the original equations and the related constraint equations. To illustrate our approach we shall focus on multi-NLS system and its two reductions (\ref{non1}), (\ref{non2}) and (\ref{non3}), (\ref{non4}).

\noindent \textbf{i)} Soliton solutions of the reduced equations (\ref{non1}), (\ref{non2}).

One-soliton solutions of the original multi-NLS equations (\ref{4NLS-a})-(\ref{4NLS-d}) are given in (\ref{sol1})-(\ref{sol4}) and the constraint equations are $p=\rho_1 q$, $s=\rho_2 r$. Recall that we have taken $p_1=q, p_2=p, q_1=r, q_2=s$, previously. The reduction equations put conditions on the parameters of the soliton solutions of the original multi-NLS equations. By Type 1 approach \cite{GurPek1}, \cite{GurPek3}, \cite{GurPek2} that is based on equating numerators and denominators separately, we obtain that in order that the soliton solutions (\ref{sol1})-(\ref{sol4}) of the original multi-NLS (\ref{4NLS-a})-(\ref{4NLS-d}) to satisfy the reduced equations (\ref{non1}), (\ref{non2}) the parameters must satisfy  the following conditions:
\begin{equation}
k_2=k_1,\quad k_4=k_3,\quad e^{\delta_2}=\rho_1e^{\delta_1},\quad e^{\delta_4}=\rho_2e^{\delta_3},
\end{equation}
yielding $\omega_2=\omega_1$, $\omega_4=\omega_3$, $\gamma_j=0$ for $j=1,2,3,4$, $M=0$, and hence one-soliton solutions of the reduced local equations (\ref{non1}), (\ref{non2}) are given by
\begin{eqnarray}\displaystyle
&q=\frac{1}{2}e^{\phi-\delta}\mathrm{sech}(\psi),\\
&r=\frac{1}{2}e^{-\phi-\delta}\mathrm{sech}(\psi),
\end{eqnarray}
where $\phi=\frac{(k_1-k_3)}{2}x+\frac{(\omega_1-\omega_3)}{2}t+\frac{(\delta_1-\delta_3)}{2}$ and $\psi=\frac{(k_1+k_3)}{2}x+\frac{(\omega_1+\omega_3)}{2}t+\frac{(\delta_1+\delta_3)}{2}+\delta$ for $\delta=\ln\Big|\frac{\sqrt{1+\rho_1\rho_2}}{k_1+k_3}\Big|$.\\

\noindent \textbf{ii)} Soliton solutions of the reduced equations (\ref{non3}), (\ref{non4}).

Using the above prescription we find the following constraints to be satisfied by the parameters of one-soliton solution (\ref{sol1})-(\ref{sol4}):
\begin{equation}
k_2=-k_1,\quad k_4=-k_3,\quad e^{\delta_2}=\rho_1e^{\delta_1},\quad e^{\delta_4}=\rho_2e^{\delta_3},
\end{equation}
yielding $\omega_2=\omega_1$, $\omega_4=\omega_3$, $\gamma_2=\gamma_1$, and $\gamma_4=\gamma_3$. Recall that here we have $\rho_1\rho_2=1$. Hence one-soliton solutions of the reduced nonlocal system (\ref{non3}), (\ref{non4}) for $(\epsilon_1,\epsilon_2)=(-1,1)$ are given by
\begin{align}\displaystyle
&q=\frac{e^{k_1x+\omega_1t+\delta_1}+\tilde{\gamma}_1e^{-k_3x+(2\omega_1+\omega_3)t+2\delta_1+\delta_3}}{1+\frac{2e^{(\omega_1+\omega_3)t+\delta_1+\delta_3}}{(k_1+k_3)^2}\cosh ((k_1+k_3)x)+\tilde{M}e^{2(\omega_1+\omega_3)t+2(\delta_1+\delta_3)}},\\
&r=\frac{e^{k_3x+\omega_3t+\delta_3}+\tilde{\gamma}_3e^{-k_1x+(\omega_1+2\omega_3)t+\delta_1+2\delta_3}}{1+\frac{2e^{(\omega_1+\omega_3)t+\delta_1+\delta_3}}{(k_1+k_3)^2}\cosh ((k_1+k_3)x)+\tilde{M}e^{2(\omega_1+\omega_3)t+2(\delta_1+\delta_3)}},
\end{align}
where
\begin{equation}\displaystyle
\tilde{\gamma}_1=\frac{2k_1}{(k_1-k_3)(k_1+k_3)^2},\, \tilde{\gamma}_3=-\frac{2k_3}{(k_1-k_3)(k_1+k_3)^2},\,\tilde{M}=-\frac{4k_1k_3}{(k_1-k_3)^2(k_1+k_3)^4}.
\end{equation}
The approach outlined above is applicable also to all kinds of other reduced equations to find any multiple soliton solutions.

\subsection{Integrability of the reduced equations}

Let a system of evolutionary type of equations for the vectorial functions $u$ and $v$ be given by
\begin{eqnarray}
&&cu_{t}=F(u,v, u_{x},v_{x}, \cdots, u_{n}, v_{n}), \label{evo1}\\
&&cv_{t}=G(u,v, u_{x},v_{x}, \cdots, u_{n}, v_{n}), \label{evo2}
\end{eqnarray}
be integrable. Here $c$ is an arbitrary constant. Then there exist two operators $R$ and $K^\star$  satisfying
\begin{equation}
\mathcal{R}_{t}= \mathcal{R}\, K^\star-K^\star \mathcal{R}
\end{equation}
where  $\mathcal{R}$ is the recursion operator and $K^\star$ is given by
\begin{equation}
K^\star= \left[\begin{array}{ll} \delta_{u} F & \delta_{u} G\cr
                        \delta_{v} F& \delta_{v} G
                        \end{array} \right]
\end{equation}
where $\delta_{u} F$ and $\delta_{v} G$ are Fr\'{e}chet derivatives of $F$ and $G$ in the directions of $u$ and $v$, respectively.

All integrable hierarchies of the systems are given by
\begin{equation}
cU_{t}=\mathcal{R}^{n}\, U_{x}, ~~~n=1,2,\cdots
\end{equation}
where $U=(u,v)^T$ and $c$ is an arbitrary constant. Let $v=k u(\epsilon_{1} t, \epsilon_{2} x)$ be a consistent reduction (local or nonlocal) of system (\ref{evo1}), (\ref{evo2}), i.e. the equation (\ref{evo2}) is not independent anymore, it can be obtained from the equation (\ref{evo1}). Here $\epsilon_{1}^2=\epsilon_{2}^2=1$ and $k$ is a constant. If the reduction is done in a consistent way the reduced system of equations
is also integrable \cite{GurPek3}. This means that the reduced system admits a recursion operator
and bi-Hamiltonian structure and the reduced system has $N$-soliton solutions. The
inverse scattering method (ISM) can also be applied. Ablowitz and Musslimani have
first found the nonlocal NLS equation from the coupled AKNS equations and solved
it by ISM \cite{AbMu1}-\cite{AbMu3}. Recursion operator of the reduced system is $R_{\epsilon}$ which is obtained from recursion operator $R$ of
the system by using the reduction. Then the integrable hierarchy of the reduced system is given by
\begin{equation}
cU^{\epsilon}_{t}=\mathcal{R}_{\epsilon}^{n}\, U^{\epsilon}_{x}, ~~~n=1,2,\cdots
\end{equation}
where $U^{\epsilon}=(u(x,t), u(\epsilon_{1} x, \epsilon_{2} t))^T$. It can be shown that the reduced recursion operator satisfies
\begin{equation}\label{reccondeps}
\mathcal{R}_{\epsilon ,t}= \mathcal{R}_{\epsilon}\, K_{\epsilon}^\star-K_{\epsilon}^\star \mathcal{R}_{\epsilon},
\end{equation}
where $K_{\epsilon}^\star$ is obtained from $K^\star$ by using the reduction. The reduced system admits also bi-Hamiltonian structure. The Hamiltonian operators of the reduced system are $J^{\epsilon}_{1}$ and $J^{\epsilon}_{2}$ which are obtained from the Hamiltonian operators $J_{1}$ and $J_{2}$ by using the reductions. The recursion operator $\mathcal{R}_{\epsilon}$ of the reduced system is given also as $\mathcal{R}_{\epsilon}=(J^{\epsilon}_{1})^{-1}\,J^{\epsilon}_{2}$. Hence we expect that the Hamiltonian operators  $J_{1}$ and $J_{2}$ satisfy the Jacobi identities (taking into account of the remarks pointed out by Achic et al. \cite{achic}).

Although we studied the integrable reductions and gave examples in  \cite{GurPek3} we shall  consider the hierarchy (\ref{hierarchy}) for $N=2$ as an example. The recursion operator for $N=2$ can be explicitly given as

{\tiny \begin{equation}
\mathcal{R}=\frac{1}{\alpha}\left( \begin{array}{cccc}
-(\partial+2q\partial^{-1}r+p\partial^{-1}s)&-q\partial^{-1}s &-2 q\partial^{-1}q &- (q\partial^{-1}p+p\partial^{-1}q) \\
- p\partial^{-1} r& -(\partial+q\partial^{-1}r+2p\partial^{-1} s) & -(p\partial^{-1}q+q\partial^{-1}p) & -2 p\partial^{-1} p\\
2 r\partial^{-1}r& (r\partial^{-1}s+s\partial^{-1}r) & (\partial+2r\partial^{-1}q+s\partial^{-1}p) &  r\partial^{-1} p\\
 (s\partial^{-1}r+r\partial^{-1}s) &2 s\partial^{-1}s &  s\partial^{-1} q &(\partial + r\partial^{-1}q+2s\partial^{-1}p)
 \end{array} \right).
\end{equation}}
Let us apply the reduction $p(x,t)=\rho_{1}\, q(\epsilon_{1}\,x, \epsilon_{2} t)=\rho_1q^{\epsilon}$ and $s(x,t)=\rho_{2}\, r(\epsilon_{1}\,x, \epsilon_{2} t)=\rho_2r^{\epsilon}$ to the hierarchy (\ref{hierarchy}) for $N=2$. Recall that for consistency we have $\rho_1\rho_2=1$ and $(\epsilon_1,\epsilon_2)=(\pm 1, 1)$. Then we have

{\tiny \begin{equation}\label{R_2eps}
\mathcal{R}_{\epsilon}=\left( \begin{array}{cccc}
-(\partial+2q\partial^{-1}r+q^{\epsilon}\partial^{-1}r^{\epsilon})&-\rho_2q\partial^{-1}r^{\epsilon} &-2 q\partial^{-1}q &- \rho_1(q\partial^{-1}q^{\epsilon}+q^{\epsilon}\partial^{-1}q) \\
- \rho_1q^{\epsilon}\partial^{-1} r& -(\partial+q\partial^{-1}r+2q^{\epsilon}\partial^{-1} r^{\epsilon}) & -\rho_1(q^{\epsilon}\partial^{-1}q+q\partial^{-1}q^{\epsilon}) & -2\rho_1^2 q^{\epsilon}\partial^{-1} q^{\epsilon}\\
2 r\partial^{-1}r& \rho_1(r\partial^{-1}r^{\epsilon}+r^{\epsilon}\partial^{-1}r) & (\partial+2r\partial^{-1}q+r^{\epsilon}\partial^{-1}q^{\epsilon}) &  \rho_1r\partial^{-1} q^{\epsilon}\\
 \rho_2(r^{\epsilon}\partial^{-1}r+r\partial^{-1}r^{\epsilon}) &2\rho_2^2 r^{\epsilon}\partial^{-1}r^{\epsilon} &  \rho_2r^{\epsilon}\partial^{-1} q &(\partial + r\partial^{-1}q+2r^{\epsilon}\partial^{-1}q^{\epsilon})
 \end{array} \right)
\end{equation}}
and
{\tiny \begin{equation}\label{Keps}
\mathcal{K}_{\epsilon}^{\star}=\left( \begin{array}{cccc}
-(\partial^2+4qr+2q^{\epsilon}r^{\epsilon})&-\rho_2qr^{\epsilon} &-2 q^2 &-2qq^{\epsilon} \\
- 2q^{\epsilon}r& -(\partial^2+4q^{\epsilon}r^{\epsilon}+2qr) & -2qq^{\epsilon} & -2 (q^{\epsilon})^2\\
2 r^2& 2rr^{\epsilon} & (\partial^2+4qr+2q^{\epsilon}r^{\epsilon}) &  2q^{\epsilon}r\\
 2rr^{\epsilon} &2(r^{\epsilon})^2 &  2qr^{\epsilon} &(\partial^2 +4q^{\epsilon}r^{\epsilon}+2qr)
 \end{array} \right).
\end{equation}}
We take t-derivative of the operator (\ref{R_2eps}) and use
\begin{align*}
&q_t=\frac{1}{c\alpha}[-q_{xx}-2(qr+q^{\epsilon}r^{\epsilon})q],\\
&q_t^{\epsilon}=\frac{1}{c\alpha}[-q_{xx}^{\epsilon}-2(qr+q^{\epsilon}r^{\epsilon})q^{\epsilon}],\\
&r_t=\frac{1}{c\alpha}[r_{xx}+2(qr+q^{\epsilon}r^{\epsilon})r],\\
&r_t^{\epsilon}=\frac{1}{c\alpha}[r_{xx}^{\epsilon}+2(qr+q^{\epsilon}r^{\epsilon})r^{\epsilon}].\\
\end{align*}
It is straightforward to see that the condition (\ref{reccondeps}) is satisfied  by the operators $\mathcal{K}_{\epsilon}^{\star}$ and  $\mathcal{R}_{\epsilon}$. This proves that  $\mathcal{R}_{\epsilon}$ is the recursion operator of the reduced system above. The only restriction comes on $\rho_1$ and $\rho_2$ as $\rho_1=\rho_2=1$.

\section{Negative multi-AKNS system}

We can easily obtain the negative hierarchy of the integrable system of equations if they admit a recursion operator \cite{GurPek4}, \cite{gur-pek02},  \cite{gur-pek0}. By using this approach negative hierarchy of any integrable system of equations is obtained from the following equations:
\begin{equation}
\mathcal{R}_N(U_t)-a\mathcal{R}_N^n (U_x)=b U_{y},~~~ n=0,1,2,\cdots,
\end{equation}
for arbitrary constants $a, b$. Here $\mathcal{R}_N$ is the recursion operator and $U=({\bf p}, {\bf q})^T$.

\vspace{0.5cm}
\noindent
For $n=0$ we have
\begin{eqnarray}\displaystyle
&&\left(R_{1}\right)^{i}_{j}\, p_t^j+\left(R_{2}\right)^{i}_{j}\, q_t^j=a p_x^j+bp_y^j,\\
&&\left(R_{3}\right)^{i}_{j}\, p_t^j+\left(R_{4}\right)^{i}_{j}\, q_t^j=a q_x^j+bq_y^j,
\end{eqnarray}
for $i=1,\cdots, N$, or equivalently
\begin{align}\displaystyle
-\frac{1}{\alpha}(p_{xt}^i+p^i\partial^{-1} (p_kq^k)_t)+p_{k}\partial^{-1} (p^i\,q^k)_t&=a p_x^i+bp_y^i,\\
\frac{1}{\alpha}(q_{xt}^i+q^i\partial^{-1} (q_kp^k)_t)+q^k\partial^{-1} (p_kq^i)_t&=a q_x^i+bq_y^i.
\end{align}

For instance if we consider $N=2$ we get explicitly
\begin{align}\displaystyle
-\frac{1}{\alpha}[p_{xt}^1+2p^1\partial^{-1}(p^1q^1)_t+p^2\partial^{-1}(p^1q^2)_t+p^1\partial^{-1}(p^2q^2)_t]&=ap_x^1+bp_y^1,\\
-\frac{1}{\alpha}[p_{xt}^2+2p^2\partial^{-1}(p^2q^2)_t+p^1\partial^{-1}(p^2q^1)_t+p^2\partial^{-1}(p^1q^1)_t]&=ap_x^2+bp_y^2,\\
\frac{1}{\alpha}[q_{xt}^1+2q^1\partial^{-1}(p^1q^1)_t+q^1\partial^{-1}(p^2q^2)_t+q^2\partial^{-1}(p^2q^1)_t]&=aq_x^1+bq_y^1,\\
\frac{1}{\alpha}[q_{xt}^2+2q^2\partial^{-1}(p^2q^2)_t+q^2\partial^{-1}(p^1q^1)_t+q^1\partial^{-1}(p^1q^2)_t]&=aq_x^2+bq_y^2.
\end{align}

\vspace{0.5cm}

\noindent For $n=1$ we have
\begin{eqnarray}\displaystyle
&&\left(R_{1}\right)^{i}_{j}\, (p_t^j-a p_x^j)+\left(R_{2}\right)^{i}_{j}\, (q^{j}_{t}-a q_x^j)=bp_y^j,\\
&&\left(R_{3}\right)^{i}_{j}\, (p_t^j-a p_x^j)+\left(R_{4}\right)^{i}_{j}\,(q^{j}_{t}-a q_x^j)=bq_y^j,
\end{eqnarray}
for $i=1,\cdots, N$. Particularly, for $N=2$ the above system becomes
\begin{align}
\displaystyle
-\frac{1}{\alpha}[p_{xt}^1+2p^1\partial^{-1}(p^1q^1)_t+p^2\partial^{-1}(p^1q^2)_t&+p^1\partial^{-1}(p^2q^2)_t]\nonumber\\
&=bp_y^1-\frac{a}{\alpha}[p_{xx}^1+2(p^1q^1+p^2q^2)p^1],\\
-\frac{1}{\alpha}[p_{xt}^2+2p^2\partial^{-1}(p^2q^2)_t+p^1\partial^{-1}(p^2q^1)_t&+p^2\partial^{-1}(p^1q^1)_t]\nonumber\\
&=bp_y^2-\frac{a}{\alpha}[p_{xx}^2+2(p^1q^1+p^2q^2)p^2],
\end{align}
\begin{align}
\frac{1}{\alpha}[q_{xt}^1+2q^1\partial^{-1}(p^1q^1)_t+q^1\partial^{-1}(p^2q^2)_t&+q^2\partial^{-1}(p^2q^1)_t]\nonumber\\
&=bq_y^1+\frac{a}{\alpha}[q_{xx}^1+2(p^1q^1+p^2q^2)q^1],\\
\frac{1}{\alpha}[q_{xt}^2+2q^2\partial^{-1}(p^2q^2)_t+q^2\partial^{-1}(p^1q^1)_t&+q^1\partial^{-1}(p^1q^2)_t]\nonumber\\
&=bq_y^2+\frac{a}{\alpha}[q_{xx}^2+2(p^1q^1+p^2q^2)q^2].
\end{align}

\vspace{0.5cm}

\noindent For $n=2$ we have
\begin{eqnarray}\displaystyle
&&\left(R_{1}\right)^{i}_{j}\, w_{1}^{j}+\left(R_{2}\right)^{i}_{j}\, w_{2}^{j}=bp_y^j,\\
&&\left(R_{3}\right)^{i}_{j}\, w_{1}^{j}+\left(R_{4}\right)^{i}_{j}\,w_{2}^{j}=bq_y^j,
\end{eqnarray}
where
\begin{eqnarray}\displaystyle
 w_{1}^{i}&=&p_t^i+\frac{1}{\alpha}\,p_{xx}^i+\frac{2}{\alpha}\,Q p^{i},\\
 w_{2}^{i}&=&q_t^i-\frac{1}{\alpha}\,q_{xx}^i-\frac{2}{\alpha}\,Q q^{i},
\end{eqnarray}
with $Q=q^kp_k$ and $i=1,\cdots, N$. For a particular value of $N$, e.g. $N=2$, the above system turns to be
\begin{align}
\displaystyle
-\frac{1}{\alpha}[p_{xt}^1+2p^1\partial^{-1}(p^1q^1)_t&+p^2\partial^{-1}(p^1q^2)_t+p^1\partial^{-1}(p^2q^2)_t]\nonumber\\
&=bp_y^1+\frac{a}{\alpha^2}[p_{xxx}^1+3(p^1q^1+p^2q^2)p_x^1+3(p_x^1q^1+p_x^2q^2)p^1],\\
-\frac{1}{\alpha}[p_{xt}^2+2p^2\partial^{-1}(p^2q^2)_t&+p^1\partial^{-1}(p^2q^1)_t+p^2\partial^{-1}(p^1q^1)_t]\nonumber\\
&=bp_y^2+\frac{a}{\alpha^2}[p_{xxx}^2+3(p^1q^1+p^2q^2)p_x^2+3(p_x^1q^1+p_x^2q^2)p^2],\\
\frac{1}{\alpha}[q_{xt}^1+2q^1\partial^{-1}(p^1q^1)_t&+q^1\partial^{-1}(p^2q^2)_t+q^2\partial^{-1}(p^2q^1)_t]\nonumber\\
&=bq_y^1+\frac{a}{\alpha^2}[q_{xxx}^1+3(p^1q^1+p^2q^2)q_x^1+3(q_x^1p^1+q_x^2p^2)q^1],\\
\frac{1}{\alpha}[q_{xt}^2+2q^2\partial^{-1}(p^2q^2)_t&+q^2\partial^{-1}(p^1q^1)_t+q^1\partial^{-1}(p^1q^2)_t]\nonumber\\
&=bq_y^2+\frac{a}{\alpha^2}[q_{xxx}^2+3(p^1q^1+p^2q^2)q_x^2+3(q_x^1p^1+q_x^2p^2)q^2].
\end{align}
As we observe from the above expressions, $(2+1)$-extensions of the multi-AKNS systems are very complicated for all $n$ and for all $N$. The Hirota bilinearization and the soliton solutions of these equations will be communicated later.

\section{Conclusion}
We presented the first two members of a multi-component AKNS hierarchy, namely multi-NLS and multi-MKdV systems. We derived Hirota bilinear forms of these systems.
We obtained one-soliton solutions of them for $N=2$ (four fields). One-soliton solutions of the multi-NLS and multi-MKdV systems for $N=2$ are the same except the dispersion relations which is a known fact for the soliton solutions of the integrable equations in the same hierarchy \cite{superpos}.

We obtained some new local and nonlocal systems of nonlinear evolutionary partial differential equations by using local and nonlocal reductions of multi-NLS and multi-MKdV equations. We gave a method to find the soliton solutions of the reduced equations from the soliton solutions of the multi-NLS and multi-MKdV equations. We also analyzed the integrability of the reduced equations. We finally constructed $(2+1)$-dimensional  negative multi-component AKNS hierarchy in general and gave examples for  $N=2$.\\

\noindent \textbf{Acknowledgments}\\

  This work is partially supported by the Scientific
and Technological Research Council of Turkey (T\"{U}B\.{I}TAK).\\

\end{document}